\documentclass[12pt]{article}

\usepackage{epsfig}
\usepackage[usenames]{color}

\def\be{\begin{equation}}
\def\ee{\end{equation}}
\newcommand{\bea}{\begin{eqnarray}}
\newcommand{\eea}{\end{eqnarray}}
\def\bd{\begin{displaymath}}
\def\ed{\end{displaymath}}

\def\rt{\rightarrow}
\definecolor{red}{rgb}{1,0,0}

\def\pa{\partial}
\newcommand{\non}{\nonumber \\}
\newcommand{\CR}{\non\cr}

\begin{document}

\vspace{18pt} \vskip 0.01in \hfill TAUP-287908 \vskip 0.01in \hfill
WIS/03/10-Feb-DPPA \vskip 0.01in \hfill {\tt hep-th/yymmnnn}

\vspace{30pt}

\begin{center}
{\bf \LARGE   Non-Perturbative Field Theory- }

{\bf
From two dimensional conformal field theory

to QCD in four dimensions }

\end{center}

\vspace{30pt}

\begin{center}
Yitzhak Frishman

\textit{{\it Department of Particle Physics and Astrophysics\\
 The Weizmann Institute of Science\\
Rehovot 76100, Israel\\[10pt]
 }}
\vspace{20pt}
Jacob Sonnenschein

\textit{{\it School of Physics and Astronomy\\ The Raymond and Beverly
Sackler Faculty of Exact Sciences\\ Tel Aviv University, Ramat Aviv
69978,
Israel\\[10pt]
 }}

\end{center}


\begin{center}
\textbf{Abstract }
\end{center}
This note is based on the summary of our book entitled
``Non-perturbative field theory- from two dimensional conformal
field theory to QCD in four dimensions",  published recently  by
{\it Cambridge University Press}. It includes 436 pages.

The book  provides a detailed description of the tool box of
non-perturbative techniques, presents applications of  them to
simplified systems, mainly of gauge dynamics in two dimensions, and
examines the lessons one can learn from those systems about four
dimensional QCD and hadron physics.

In particular the book deals with  conformal invariance,
integrability, bosonization,  large N, solitons in two dimensions and monopoles and instantons in four dimensions,
confinement versus screening and finally the hadronic spectrum and
scattering.

We also attach the table of contents and the list of references of
the book.

{\bf We would be grateful for any comments or suggestions related to
the material in the book. These may be incorporated in a possible
future edition. They may be sent via the e-mails below.}


\vspace{4pt} {\small \noindent

 }  \vfill
\vskip 5.mm
 \hrule width 5.cm
\vskip 2.mm {\small \noindent yitzhak.frishman@weizmann.ac.il\\
cobi@post.tau.ac.il}

\thispagestyle{empty}

\eject

\setcounter{page}{1}

\def\VEV#1{\left\langle #1\right\rangle}

 \section{General}
Relativistic Quantum Field Theory has been very successful in
describing strong, electromagnetic and weak interactions, in the
region of small couplings by using perturbation theory, within the
framework of the Standard Model.

However, the region of strong coupling, like the hadronic spectrum
and various scattering phenomena of hadrons within QCD, is still
largely unsolved.

 A large variety of methods have been  used  to address this question,
 including    QCD sum rules,    lattice gauge  simulations, light cone quantization, low energy effective
 Lagrangians like the Skyrme model and chiral Lagrangians,
 large $N$ approximation, techniques of conformal invariance, integrable model approach,
 supersymmetric models, string theory approach
 etc.
 In spite of this  major  effort the gap between the phenomenology and
 the basic  theory has been  only partially bridged, and the problem is still open.

The goals of this book are to provide a detailed description of the
tool box of non-perturbative techniques, to apply them on simplified
systems, mainly  of gauge dynamics in two dimensions, and to examine
the lessons one can learn from those systems about four dimensional
QCD and hadron physics.

The study of  two dimensional models in order  to improve the understanding of
 four dimensional  physical systems
  was found to be fruitful.
 This can be achieved following two different approaches. In the first one applies  non-perturbative methods on the simpler two dimensional model, extract the physical behavior and extrapolate it to four dimensions. The second is based on gaining insight on the methods by first applying them first in two dimensions, then apply the analogous methods on the four dimensional system and deduce certain physical properties of it.

   This follows two directions, one is the utilization of
  non-perturbative methods  on simpler setups and the second is extracting the physical behavior of
  hadrons in one space dimension.

Obviously,
physics in two dimensions is simpler than that of the real world since the
underlying manifold is  simpler and since  the number of
degrees of freedom of
each field is smaller. There are some additional
simplifying features in two
dimensional physics. In one space dimension
there is no  rotation symmetry and no angular momentum. The light
cone is disconnected and  is composed of
 left moving and right moving branches.
 Therefore, massless particles are either
 on one branch or the other.
 These two  properties are the  basic building
blocks of the idea of transmutation between
systems of different statistics.
Also, the ultra-violet behavior is more convergent in two dimensions,
making for instance $QCD_2$ a superconvergent theory.

In this summary note we go over several  notions, concepts and
methods with an emphasis on the comparison between the two and four
dimensional worlds and on what one can deduce about the latter from the
former. In particular we deal with conformal invariance,
integrability, bosonization, large N, solitons in two dimensions and monopoles and instantons in four dimensions,
confinement versus screening, and finally the hadronic spectrum and
scattering. We end  this note with an brief outlook which includes  several comments on (i) further
progress in the application of the methods discussed in the book, (ii) applications to other domains and developments
in gauge dynamics due to other methods.
The table of content of the book and the bibliography of the book are added as appendices.

\section{Conformal Invariance}
From the onset there is a very dramatic difference between conformal
invariance  in two and four dimensions. The former is characterized
by an infinite dimensional algebra, the Virasoro algebra, whereas
the latter is associated with the finite dimensional algebra of
$SO(4,2)$. This basic difference stems from the fact that whereas
the conformal transformations in four dimensions are global, in two
dimensions the parameters of conformal transformations are
holomorphic functions (or anti-holomorphic). Nevertheless there are
several features of conformal invariance which are common to the two
cases. We will now compare various aspects of conformal invariance
in two and four dimensions.
\begin{itemize}
\item
The notion of  a primary field and correspondingly a highest weight state is used both in
two dimensional conformal field theories as well as for the four dimensional collinear algebra.
\footnote{ Define the  lightcone coordinate $x_-$  via $x^\mu = x_- n_+^\mu + x_+n_-^\mu + x_T^\mu$
where $n_+^\mu {n_+}_\mu=n_-^\mu {n_-}_\mu=0$ $ n_+^\mu {n_-}_\mu =1$. The collinear group which is an $SL(2,R)$ group, is defined by the following three transformations $x_-\rt x_-+ a_-$, $ x_-\rt ax_-$ and $ x_-\rt \frac{x_-}{1+2\tilde a x_-}$ where $\tilde a^\mu= a n^\mu$}
It is  expressed  for the former as
\be L_0
[\phi(0)|0>]= h [|\phi(0)|0>] \qquad L_n [\phi(0)|0>] =0, \qquad n>0 \ee
 and for the latter
\be
L_0 [\Phi(0)|0>]= j [|\Phi(0)|0>] \qquad L_- [\Phi(0)|0>] =0,
\ee
The difference is of course the  infinite set of annihilation operators $L_n$ versus the single annihilation operator $L_-$ in four dimensions.
\item
The COPE, the conformal operator product expansion, has a compact
form in two dimensional CFT \be
  \O_i(z,\bar z ) \O_j(w,\bar w) \sim \sum_k C_{ijk}(z-w)^{h_k-h_i-h_j}
(\bar z-\bar w)^{\bar h_k-\bar h_i-\bar h_j}\O_k(w,\bar w) \ee where $
C_{ijk}$ are  the  {\it product  coefficients}
while in four dimensions it reads
\bea A(x) B(0) &=& \sum_{n=0}^\infty  C_{n}\left(\frac{1}{x^2}
\right )^{1/2(t_A+t_B- t_n)} \frac{x_-^{n+s_1+s_2-s_A-s_B
}}{B(j_A-j_B+j_n, j_B-j_A+j_n)} \CR
 &\times& \int_0^1 du u ^{(j_A-j_B+j_n-1)}(1-u)^{(j_B-j_A+j_n-1)}{\cal O}_{n}^{j_1,j_2} (u x_-)
\eea where the definitions of the various quantities are in Chapter
17 of the book. Again there is a striking difference between the
simple formula in two dimensions and the complicated one in four
dimensions.
\item
As an example let's compare the OPE of two currents. As is described
in  Chapter 3,  the expression in two dimensions
reads \be J^a(z)J^b(w)= {k\delta^{ab}\over (z-w)^2} + i {f^{ab}_c
J^c(w)\over (z-w)} +finite\  terms \ee for any non-abelian group,
and in particular for the abelian case the second term on the RHS is
missing. For comparison the OPE of the transverse components of the
electromagnetic  currents given in Chapter 17 takes the form \bea &
J^T(x)J^T(0)\sim \CR & \sum_{n=0}^{\infty} C_n \left(
\frac{1}{x^2}\right )^{(6-t_n)/2}
(-ix_-)^{n+1}\frac{\Gamma(2j_n)}{\Gamma(j_n)\Gamma(j_n)}\int_0^1 du
[u(1-u)]^{j_n-1}{\cal Q}^{1,1}_n(ux_-)\CR \eea
\item
The conformal Ward identity associated with the dilatation operator
in four dimensions
 \be
\sum_i^N (l_\phi+ \gamma(g^*) + x_i \pa_i) < T
\phi(x_1)...\phi(x_N)> = 0 \ee where $l_\phi$ is the canonical
dimension and  $\gamma(g^*)$ is the anomalous dimension,  seems very
similar to the one in two dimensions \be
\sum_i(z_i\pa_i+h_i)<0|\phi_1(z_1,\bar z_1)...\phi_n(z_n,\bar
z_n)|0> =0\ee In both cases one has to determine the full quantum
conformal dimensions of the various operators. However, as is shown
in Chapter 2, in certain CFT models, like the unitary minimal
models, there are powerful tools based on unitarity which enable us
to determine exactly the dimensions $h_i$ of all the primary
operators and hence all the operators of the model. On the other
hand, it is a non-trivial task to determine the anomalous dimensions
in other models in two dimension, and of course of  four dimensional
operators. In certain supersymmetric theories there are operators
whose dimension is protected \footnote {Of course besides the usual
operators with no anomalous dimensions, like conserved currents,
energy-momentum tensor, and the like}, but generically one has to
use perturbative calculations to determine the anomalous dimensions
of gauge theories to a given order in the coupling constant.

Using the Ward identity one can extract the form of the two point
function of operators of spin $s$ in four dimensions. It is given by
\be < \phi(x_1) \phi(x_2) > = N_2( g^*)(\mu^*)^ {-2\gamma(g^*)}
\left [ \frac{1}{(x_1-x_2)^2}\right ]^{ l_\phi+ \gamma(g^*)}\left (
\frac{(x_1-x_2)_+}{(x_1-x_2)_-} \right )^s \ee The corresponding two
point function in two dimensions, which depends only on the
conformal dimension of the operator $h$, reads \be G_2(z_1,\bar z_1,
z_2,\bar z_2)\equiv
  <0|\phi_1(z_1,\bar z_1)\phi_1(z_2,\bar z_2)|0>
 = {c_2\over (z_1-z_2)^{2h_1}(\bar z_1-\bar z_2)^{2\bar h_1}}
\ee
\item
As for higher point functions, it is shown in Chapter 2 that one
can use the local Ward identities together with Virasoro null
vectors to write down  a partial differential equations that determine the correlators. The result for
a four point function was later used to determine the four point
function of the Ising model.
\item
Two dimensional conformal field
theories are further invariant under affine Lie algebra
transformations, and as is  shown in Chapter 3 those can be
combined with null vectors to derive the so called
Knizhnik-Zamolodchikov equations, which can be  used to solve for
the four-point function of the $SU(N)$ WZW model ( see  Chapter 4). This
type of differential equations that fully determine correlation
functions are obviously absent in four dimensional interacting
conformal field theories.

\end{itemize}

\section{Integrability}
 Integrability is discussed in Chapter 5
in the context of two dimensional models and in Chapter 18 in four
dimensional gauge theories. For systems with a finite number of
degrees of freedom, like spin chain models, there is a finite number
of conserved charges, equal of course to the number of degrees of
freedom. For integrable field theories there is an infinite
countable number of conserved charges. Furthermore, the scattering
processes of those models always involve a conservation of the
number of particles.

In two dimensions we have encountered continuous  integrable models like the sine-Gordon model as well as
 discretized ones like the XXX spin chain model.  The integrable sectors of gauge dynamical systems
are based on identifying an exact map between certain properties
 of the  systems and  a  spin chain structure.
In two dimensions the spin chain models follow  from a
discretization of the space coordinate, by placing a spin variable
on each site that  can take several values and imposing periodicity.
In the four dimensional ${\cal N}=4$  super YM theory the spin chain
corresponds to a trace of field operators and in the process of
high-energy scattering it is a ``chain" of reggeized gluons
exchanged in the t-channel of a scattering process.
 A summary of the comparison among the basic two dimensional spin chain, the ``spin chains"
 associated with the planar ${\cal N}=4$ SYM,  and the high energy scattering in QCD, is given in table 1.
 \begin{table}
\centering

\begin{tabular}{|c|c|c|c|}
\hline\hline
Spin chain& Planar    & High energy   \\
& ${\cal N}=4$ SYM   &  scattering in QCD  \\
\hline
Cyclic  & Single trace  & Reggeized guons \\
 spin chain &  operator &  in t-channel \\\hline
Spin  & Field operator & $SL(2)$ spin  \\
 at a site & &   \\\hline
Number of  & Number of  & Number of    \\
 sites &  operators &   gluons  \\\hline
Hamiltonian & Anomalous dilatation  & $H_{BFKL}$ \\
 &  operator &  \\\hline
Energy   & Anomalous  dimension  & $\sim \frac{1}{\lambda}\frac{log {\cal A}}{log s}$ \\
 eigenvalue  &   $g^{-2}\delta {\cal D}$ &  \\\hline
evolution  &  dilatation & the total rapidity  \\
 time & variable &  $log s$ \\\hline
 Zero momentum  $U=1$& Cyclicity constraint &  \\\hline
\end{tabular}\label{tableint}\caption{spin chain structure of the two dimensional model and the four dimensional gauge systems of ${\cal N}=4$ SYM and of high-energy  behavior of scattering amplitudes in QCD. }
\end{table}.
A powerful method to solve all these spin chain models is the use of
the algebraic Bethe ansatz. This is  discussed in details for the
the $XXX_{1/2}$ model in Chapter 5. The solutions of the energy
eigenvalues needed for the high energy scattering process was based
on generalizing this method to the case of spin $s$ Heisenberg model
and for the ${\cal N}=4$ to the case of an $SO(6)$ invariance.

There is one conceptual difference between the spin chains of the
two dimensional models and those associated with the ${\cal N}=4$
SYM in  four dimensions. In the former  the models are non
conformal, involving a scale, and hence also with particles and an
S-matrix.
 The integrable sectors of four dimensional gauge
 theories, however,  are conformal invariant.

The study of integrable models in two dimensions is quite mature,
whereas the application of integrability to four dimensional systems
is  at an infant stage. The concepts of multi-local charges and of
quantum groups, discussed in Chapter 5, have been applied only
slightly to gauge dynamical systems in four dimensions.

\section{Bosonization}

Bosonization is the formulation of fermionic systems   in terms of
bosonic variables and fermionization is just the opposite process.
The detailed discussion is in Chapter 6.  The study of  bosonized
physical systems offers several advantages:\hfill\break (1) It is
usually easier to deal with commuting fields rather than
anti-commuting ones.\hfill\break (2) In certain examples, like the
Thirring model, the fermionic strong coupling regime turns into the
weak coupling one in its bosonic version, the Sine-Gordon
 model.\hfill\break
 (3) The non-abelian
bosonization,  especially in the product scheme, offers a separation
between colored and flavored degrees of freedom, which is very
convenient for analyzing the low lying spectrum.\hfill\break (4)
Baryons composed of $N_C$ quarks are a many-body problem in the
fermion language, while simple solitons in the boson language.
\hfill\break (5)  One loop fermionic computations involving the
currents turn into tree level consideration in the bosonized
version. The best known example of the latter are the chiral (or
axial) anomalies.

In four dimensions, spin is obviously non-trivial and one cannot
constitute generically a bosonization equivalence. However, in
certain circumstances a four dimensional system can be described
approximately by fields that depend only on the time and the radial
direction. In those cases one can apply the bosonization technique.
Examples of such scenarios are monopole induced proton decay,
 and fractional charges induced on
monopoles by light fermions. In these cases the relevant degrees of
freedom are in an s-wave and hence taken to depend only on the time
and the radial direction. This enables one to use the corresponding
bosonized field. There is a slight difference with two dimensions,
as the radial coordinate goes from zero to infinity, so "half" a
line. Appropriate boundary conditions enable to use a reflection, so
to extend to a full line.

\section{Topological field configurations}
\begin{itemize}
\item
The topological charges in any dimensions are conserved regardless
of the equations of motion of the corresponding systems. In two
dimensions it is very easy to write down a current which is
conserved without the use of the equations of motion. This is
referred to as a topological conservation. Consider a scalar field
$\phi$ or its non-ablelian analog  which is expressed in terms of a group element
$g\in G$ of a non-abelian group $G$,
 then the following currents are
abelian and non-abelian conserved currents \be J_\mu =
\epsilon_{\mu\nu}\pa^\nu \phi \qquad  J_\mu =
\epsilon_{\mu\nu}g^{-1}\pa^\nu g \ee

Recall that for a system that admits, for instance in abelian case,
also a current $J_\mu = \pa_\mu\phi$ that is conserved upon the use
of the equations of motion,  one can then replace the two currents
with  left and right conserved currents  $J_\pm=\pa_\pm \phi$ or
$J=\pa\phi$ and $\bar J=\bar\pa\phi$ using complex coordinates. The charge associated with the
topological conserved current is given by \be Q_{top}=\int dx
\phi'=[\phi(t,+\infty)-\phi(t,-\infty)]\equiv \phi_+-\phi_- \ee
where the space dimension is taken to be ${\cal R}$. For a
compactified space dimension, namely an $S^1$ this charge vanishes,
except for cases where the field is actually an angle variable, in
which case the charge is $2\pi$. The latter appears in the case of
U(1) gauge theory in two dimensions, where there is a winding
number.
\item
Obviously one cannot have such topologically conserved currents and
charges  in four dimensions. However, for theories that are
invariant under a
 non-abelian group, one can construct also in four dimensions a topological current and charge, like
 for the cases of Skyrmions, magnetic monopoles and instantons.
 For the Skyrmions  the topological current is given by
 \be\label{Jskyrme}
 J^\mu_{skyre}= \frac{i\epsilon^{\mu\nu\rho\sigma}}{24\pi^2} Tr[L_\nu L_\rho L_\sigma]
 \ee
where $L_\mu= U^\dagger\pa_\mu U$ with $U\in SU(N_f)$
\item
The topological charges, for compact spaces, are the winding numbers
of the corresponding topological configurations. For a compact one
space dimension, we have the map of $S^1\rightarrow S^1$ related to
the homotopy group $\pi_1(S^1)$. In two space dimensions, the
windings are associated with the map $S_2\rightarrow S_2^{G/H}$, as
for the magnetic monopoles. For three space dimensions, it is
$S^3\rightarrow S^3$ for the Skyrmions at $N_f=2$, and the
non-abelian instantons for the gauge group $SU(2)$. The topological
data of the various models is summarized in table 2.

\item
According to Derrick's theorem, for a theory of a scalar field with
an ordinary kinetic term with two derivatives, and any local
potential at $D\geq 2$, the only non-singular time-independent
solutions of finite energy are the vacua. However, as is described
in Chapters 20-22, there are solitons in the form of Skyrmions and
monopoles and instantons. Those configurations bypass Derrick's
theorem by introducing higher derivative terms or including
non-abelian gauge fields.
\item
As is emphasized in Chapter 20, the extraction of the baryonic
properties in the Skyrme model is very similar to the one for
baryons in the bosonized theory in two dimensions. Unlike the latter
which is exact in the strong coupling limit, one cannot derive the
former starting from the underlying theory. Another major difference
between the two models is of course the existence of angular
momentum only in the four dimensional case.

\begin{table}
\centering
\caption{Topological  classical field configurations in two and four dimensions}
\begin{tabular}{|c|c|c|c|}
\hline
classical    & dim. & map  & topological    \\
 field   &  &    &  current   \\
\hline soliton  & two &   & $ \epsilon^{\mu\nu}\pa_\nu \phi$
\\\hline baryon  & two & $S^1\rightarrow S^1$ & $\epsilon^{\mu\nu}Tr[g^{-1}\pa_\nu g]$; \  $g\in U(N_f)$
\\\hline Skyrmion  & four & $S^3\rightarrow S^3$  &
$\frac{i\epsilon^{\mu\nu\rho\sigma}}{24\pi^2} Tr[L_\nu L_\rho
L_\sigma] $ \\\hline monopole  & four & $ S^2_{space} \rightarrow
S^2_{G/H}$ &
$\frac{1}{8\pi}\epsilon_{\mu\nu\rho\sigma}\epsilon^{abc}
\pa^\nu\hat\Phi^a \pa^\rho\hat\Phi^b \pa^\sigma\hat\Phi^c $ \\\hline
instanton  & four & $S^3_s\rightarrow S^3_g$ &
$\frac{i\epsilon^{\mu\nu\rho\sigma}}{16\pi^2} Tr[A_\nu \pa_\rho
A_\sigma + \frac{2}{3} A_\nu A_\rho A_\sigma] $
  \\\hline
\end{tabular}\label{table23.4}
\end{table}
\item
A non-trivial task associated with topological configurations is the
construction of configurations that carry multipole topological
charge, for instance a multi-baryon state both of the bosonized
$QCD_2$ as well as of the Skyrme model, a multi-monopole solution
and a multi-intanton solution. For the two dimensional baryons (see
Chapter 13) the construction is a straightforward generalization of
the the configuration of baryon number one. For the multi-monopoles
solutions the book describes Nahm's construction, and for the
multi-intantons the ADHM construction. These constructions, which
are in fact related, are much more complicated than that for the two
dimensional muti-baryons.
\item
A  very important phenomenon that occurs in both two and four
dimensions is the strong-weak duality, and the duality between a
soliton and an elementary field. In two dimensions we have
encountered this duality in the relation between the Thirring model
and the sine-Gordon model, where the coupling of the latter $\beta$
is related to that of the former $g$ as (Chapter 6) \be
\frac{\beta^2}{4\pi} =\frac{1}{1+ \frac{g}{\pi}} \ee This also
relates the elementary fermion field of the Thirring model with the
soliton of the sine-Gordon model. In particular for $g=0$
corresponding to $\beta^2=4\pi$, the Thirring model describes a free
Dirac fermion, while the soliton of the corresponding sine-Gordon
theory is the same fermion in its bosonization disguise.
 An analog in four dimensions is the Olive-Montonen duality discussed in Chapter 21, which relates
  the electric charge $e$ with the magnetic one $e_M=\frac{4\pi}{e}$, where the former is carried by
  the elementary states $W^{\pm}$ and  the latter by the magnetic monopoles.
\end{itemize}
\section{Confinement versus screening}
The naive intuition tells us that dynamical quarks in the
fundamental representation can screen external sources in the
fundamental representation, dynamical adjoint quarks can screen
adjoint sources, but that dynamical adjoint cannot screen
fundamentals. It turns out that in two dimensions this is not the
case, and massless adjoint quarks can screen an external source in
the fundamental representation. Moreover any massless dynamical
field will necessarily be in the screening phase. The argument is
that in all cases  considered in the book  we have found that the string
tension is proportional to the mass of the dynamical quarks \be
\sigma\sim m g \ee where $m$ is the mass of the quark and $g$ is the
gauge coupling, and hence for the massless case it vanishes. This
is shown in Chapter 14 based on performing a chiral rotation that
enables us to eliminate the external sources and compute the string
tension as the difference between the Hamiltonian of the system with
the external sources and the one without them namely \be\sigma =
<H>-<H_0>\ee

It seems as if the situation in two dimensions is very different
than in four dimensions. From the onset there is a dramatic
difference between two and four dimensions relating to the concept
of confining theory. In two dimensions both the coulomb (abelian)
potential and the non-abelian one are linear with the separation
distance  $L$ whereas obviously the coulomb potential between two
particles behaves like $1/L$ while the confining one is linear with
$L$. However, that does not explain the difference between two and
four dimensions, it merely means that in two dimensions the coulomb
and confining potentials behave in the same manner. The
determination of the string tension in two dimensions cannot be
repeated in four dimensions. The reason is that in the latter case
the anomaly is not linear in the gauge fields and thus one cannot use
the chiral rotation to eliminate the external quark anti-quark pair.
That does not imply that  the situation in four dimensions differs
from the two dimensional one, it just means that one has to use
different methods to compute the string tension in four dimensions.

What are the four dimensional systems that  might resemble the two dimensional case
of dynamical adjoint matter and external fundamental quarks? A system with  external quarks in the
fundamental representation in the context of pure YM theory seems a possible analog since the
dynamical fields, the gluons, are in the adjoint representation, though they are vector fields and not fermions.
 An alternative is the ${\cal N}=1$ SYM where in addition to  the gluons  there are also  gluinos which are
 majorana fermions in the adjoint representation. Both these cases should correspond to the massless adjoint
 case in two dimensions. The latter admits a screening behavior whereas the four dimensional models seem
 to be in the confining phase. This statement is supported     by several  different types of calculations
 in particular for the non supersymmetric case this behavior is found in  lattice simulations.

At this point we cannot provide a satisfactory intuitive explanation why the behavior in two and four dimensions is so different. There is also no simple picture of how the massless adjoint dynamical quarks in two dimensions are able to screen external charges in the fundamental representation.

It is worth mentioning that there is ample evidence that four
dimensional hadronic physics  is well described by  a string theory.
This  is based for instance on realizing that mesons and baryons in
nature admit Regge trajectory  behavior which is an indication of a
stringy nature. Any string theory is by definition a two dimensional
theory and hence a very basic relation between four dimensional
hadron physics and two dimensional physics.

 In addition to the ordinary string tension which relates to the potential between a quark and anti-quark in the fundamental representation, one defines the $k$ string that connects a set of $k$ quarks with a set of $k$ anti-quarks.
This object has been examined in four dimensional YM as well as four
dimensional ${\cal N}=1$ SYM. These two cases seem to be the analog
of the two dimensional QCD theory with adjoint quarks and with
external quarks in a representation that is characterized by $k$
boxes in the Young tableau description. In Chapter 14 we present
 an expression for the  string tension as a function of the
representation  of the external and dynamical quarks and in
particular for dynamical adjoint fermions and external quarks in the
$k$ representation. If there is any correspondence between the four
dimensional adjoint matter field  and the two dimensional adjoint
quarks it must be with massive adjoint quarks since for the massless
case, as was mentioned above, the two dimensional string tension
vanishes whereas  the four dimensional one does not. Thus one may
consider a correspondence for a softly broken ${\cal N}=1$ case
where the gluinos are massive.

In two dimensions for the pure YM case we found that the string
tension behaves like $\sigma \sim g^2 k_{ext}^2$ while a Wilson line
calculation yields  $\sigma \sim g^2 C_2(R)$, where $C_2(R)$ is the
second Casimir operator in the $R$ representation of the external
quarks. For the QCD case of  general $k$  external charges and
adjoint dynamical quarks we found \be \sigma_k^{2d} \sim
\sin^2(\frac{\pi k}{N}) \ee whereas in four dimensions it is
believed that for general $k$, the string tension either follows a
Casimir law or a sinusoidal rule \be \sigma^{cas}_k \sim
\frac{k(N-k)}{N} \qquad \sigma_k^{sin}\sim \sin(\frac{\pi k}{N}) \ee

As expected all these expressions are invariant under $k\rightarrow N-k$  which corresponds for antisymmetric representations replacing a quark with an anti-quark.

\section{Hadronic phenomenology of two dimensions \\ versus four dimensions}
 $QCD_2$ was addressed first in the fermionic formulation in the
seminal work 't Hooft  where  the mesonic spectrum in the large
$N_C$ limit was determined. In the book we have  presented three additional approaches to the
hadronic spectra in two dimensions:(i)   the currentization method
for massless quarks for the entire plane of $N_C$ and $N_f$,(ii) the
DLCQ approach to extract the mesonic spectrum for the case of
fundamental as well as adjoint quarks and finally (iii) the bosonized
formulation in the strong coupling limit to determine the baryonic
spectrum. As for the four dimensional hadronic spectrum we described in the book
the use of the large $N_C$ planar limit and  the analysis of the
baryonic world using the Skyrme model. It is worth mentioning again
that whereas in the four dimensional case the Skyrme approach  is
only an approximated model derived by an ``educational guess", in
two dimension the action in the strong coupling regime is exact.

 \subsection{Mesons}

 As was just mentioned the two dimensional mesonic spectrum was extracted  using
 the large $N_C$ approximation in the fermionic formulation for $N_f=1$ ('t Hooft model),
 also by using the currentization for massless quarks and the DLCQ approach that
 can be applied for both the cases of quarks in the fundamental representation and the adjoint representation.
 For the particular region of $N_C>>N_f$ and $m=0$, the fermionic large $N_c$  and the currentization
 treatments yielded identical results. In fact this result is achieved also using the DLCQ method for
 adjoint fermions upon a truncation to a single parton and replacing $g^2$ with $2g^2$ (see
 Chapter 12).
 For massive fundamental quarks the DLCQ results match very nicely those of lattice  simulations
 and the large $N_c$ calculations, as can be seen from figures 12.1 and 12.2 in the book.

In all these methods  the corresponding  equations do not admit  exact analytic solutions for the whole range of parameters and thus   one has to resort to numerical solutions, however, in certain domains one can determine the analytic behavior of the wavefunctions and masses.

The spectrum of  mesons in two dimensions is characterized by the
dependence of the meson masses $M_{mes}$ on the gauge coupling $g$,
the number of colors $N_c$, the number of flavors $N_f$, the quark
mass $m_q$, and the excitation number $n$. In four dimensional $QCD$
the meson spectra  depend on the same parameters apart from the fact
that $\Lambda_{QCD}$ the $QCD$ scale is replacing the two
dimensional gauge coupling. The following lines summarize the
properties of the spectrum:
\begin{itemize}

  \item
  The highly excited states $n>>1$, where $n$ is the excitation number, are characterized by

  \be M_{mes}^2\sim \pi g^2 N_c n
   \ee
    This seems to fit the behaviors of mesons in nature. This behavior is referred to as a
   Regge trajectory and it follows easily from a bosonic string model of the meson.
    Following this analogy, the role of the string tension in two dimensional model is played by $g^2 N_c$.  This seems to be in contradiction with the statement that the string tension is proportional to $m_q g$.

      It is very difficult to derive the Regge trajectory behavior from direct calculations in four dimensional QCD.
\item
The opposite limit of  low lying states and in particular the ground state can be deduced
 in the limit of  large quark masses, namely $m_q>>g$  and small quark masses $g>> m_q$. For the ground state in the former limit we find
\be M_{mes}^0 \cong m_{_1} + m_{_2} \ee where $m_{i}$ are the masses
of the quark and anti-quark. In the opposite limit of $m_q<<g$ \be
(M_{mes}^0)^2 \cong \frac {\pi}{3}\sqrt{\frac{g^2N_c}{\pi}} (m_1 +
m_2) \ee For the special case of massless quarks we find a massless
meson. This is very reminiscent of the four dimensional picture. For
the massless case this should compare with the massless pions and
for small masses this is similar to the pseudo-Goldstone boson
relation where \be m_\pi^2 \sim \frac{<\bar \psi \psi>}{f_\pi^2}
(m_1 + m_2) \ee

\item
The 't Hooft model cannot be used to explore the dependence on $N_f$
the number of flavors. This can be done from the 't Hooft like
equations derived in Chapter 11. It was found out that for the first
massive state there is a  linear dependence of the meson mass
squared  on $N_f$ \be M^2_{mes}\sim N_f \ee We are not aware of a
similar behavior of the mesons in four dimensions.
\item
The 't Hooft model provides the solution of the meson spectrum in
the planar limit in two dimensions. The planar, namely large $N_c$
limit, in four dimensions is too complicated to be similarly solved.
As explained in Chapter 19 one can extract the scaling dependence in
$N_c$ of certain hadronic properties like the mass the size and
scattering amplitude, but the full determination of the hadronic
spectrum and scattering is still unresolved. A tremendous progress
has been made in the understanding of the supersymmetric theory of
${\cal N}=4$ partly by demonstrating that certain sectors of it can
be described by integrable spin chain models (see Chapter 18).
\item
As is demonstrated in Chapter 12 the DLCQ  method has been found
very effective to address the spectrum of mesons of two dimensional
QCD. This raises the question of whether one can use the DLCQ method
to handle the spectrum of  four dimensional  QCD. This task is
clearly much more difficult. En route to the extraction of the
hadronic spectrum of $QCD_4$ an easier system has been analyzed. It
is that of the collinear QCD (see Chapter 17) where in the
Hamiltonian of the system one drops off all interaction terms that
depend on the transverse momenta. In this effective two dimensional
setup the transverse degrees of freedom  of the gluon are retained
in the form of two scalar fields. This system which was not
described in the book has actually been solved  and
complete bound and continuum spectrum were extracted as well as the
Fock space wavefunctions.
\end{itemize}

\subsection{Baryons}

Chapter 13  of the book  describes  the spectrum of baryons in
multiflavor two dimensional QCD in the strong coupling limit
$\frac{m_q}{e_c}\rightarrow 0$. The four dimensional  baryonic
spectrum is discussed in the large $N_C$ limit in Chapter 19 and
using the Skyrme model approach in  Chapter 20. We would like now to
compare these spectra and to investigate the possibility of
predicting four dimensional baryonic properties from the simpler two
dimensional model. In the former case the mass is a function of the
QCD scale $\Lambda_{QCD}$, the number of colors $N_C$ and the number
of flavors $N_f$, and in the latter it is a function of $e_c, N_C$
and $N_f$. Thus it seems that the dimensionful gauge coupling in two
dimensions is the analog of $\Lambda_{QCD}$ in four dimensions.
\begin{itemize}

\item
In two dimensions, in the strong coupling limit, the mass of the baryon  was found to be
\be
E=4m \sqrt{{2N_C \over \pi}} + {m \sqrt 2}
\sqrt{{\left({\pi\over N_C}\right)}^3} \left[
C_2 - N_C^2 {(N_F-1)\over 2N_F} \right] \qquad  \label{mishE} \ee
 where the classical mass $m$ is given by
 \be m =[N_Ccm_q({e_c\sqrt{N_F}\over
\sqrt{2\pi}})^{\Delta_{C}} ]^{1\over 1+\Delta_{C}}  \ee  with
$\Delta_C ={N_C^2-1\over
N_C(N_C+N_F)}$.
Due to the fact that in two dimensions there is no spin, the structure of the spectrum with respect to the flavor group is obviously different in two and four dimensions. For instance the lowest allowed state for $N_C=N_f=3$ is in two dimensions the totally symmetric representation {\bf 10}, where as it is the mixed representation {\bf 8} in four dimensions.
\item
Let us discuss now the scaling with  $N_C$ in the large $N_C$ limit.
In two dimensions the classical term behaves like $N_C$, while the quantum correction
like $1$.  The classical result is in accordance with the result,
derived when the large $N$ expansion is applied to the baryonic
system (Chapter 19), and with the Skyrmion result (Chapter 20).
However, whereas in two dimensions the quantum correction behaves
like $N_c^0$ namely suppressed by a factor of $\frac{1}{N_c}$, in
four dimensions it behaves like $\frac{1}{N_c}$ namely a suppression
of $\frac{1}{N^2_c}$. This is summarized in table 3.
\begin{table}
\centering
\caption{Scaling of Baryon masses with $N_C$ in two and four dimensions}
\begin{tabular}{|c|c|c|}
\hline
  &  two dimensions& four dimensions   \\
\hline
Classical baryon mass &  $N_C$ &  $N_C$ \\ \hline
Quantum correction &  $N_C^0$ &  $N_C^{-1}$ \\ \hline
\end{tabular}\label{table23.2}
\end{table}

\item
In terms of the dependence on the number of flavors, it is
interesting to note that both in two dimensions and in four
dimensions, the contribution to the mass due to the quantum
fluctuations is proportional to the second Casimir operator
associated with the representation  of the baryonic state under the
$SU(N_f)$ flavor group \footnote{compare (\ref{mishE}) with (68) of Chapter
20}.
\item
Another property of the baryonic spectrum that can be compared
between the two and four dimensional cases is the flavor content of
the various states. In Chapter 13 we presented the $\bar u u, \bar d
d$ and $\bar s s $ content for the $\Delta^+$ and $\Delta^{++}$
states. Recall that in the two dimensional model for $N_C=N_f=3$ we
do not have a state in the  {\bf 8} representation but only in the
{\bf 10} so strictly speaking there is no exact analog of the
proton. Instead we take the charge =+1 $\Delta^+$ as the two
dimensional analog of the proton. In the Skyrme model one can
compute in a similar manner the flavor content of the four
dimensional baryons. The two and four dimensional states compare as
is summarized in table 4.
\begin{table}
\centering
\caption{flavor content of two dimensional and four dimensional baryons}
\begin{tabular}{|c |c c|c c|}
\hline
  &  two dimensions& & four dimensions &  \\
  &  state & value & state & value \\
\hline
$\VEV{\bar{u} u}$ & ${\Delta^+}$ & ${1\over 2}$& ${p}$& ${ 2\over 5}$\\
 \hline
 $\VEV{\bar{d} d}$ & ${\Delta^+}$ & ${1\over 3}$& ${p}$& ${ 11\over 30}$\\
 \hline
 $\VEV{\bar{s} s}$ & ${\Delta^+}$ & ${1\over 6}$& ${p}$& ${ 7\over 30}$\\
 \hline
 $\VEV{\bar{s} s}$ & $\Delta^{++}$ & ${1\over 6}$& ${\Delta}$& ${ 7\over 24}$\\
 \hline
 $\VEV{\bar{s} s}$ &  & & ${\Omega^-}$& ${ 5\over 24}$\\
 \hline
\end{tabular}\label{tab23.4}
\end{table}

\end{itemize}

\section{Outlook}
We can imagine future developments associated with the topics covered in the book in three different directions: Further progress in the application of the methods discussed in the book  to unravel the mysteries of gauge dynamics in nature, applications of the methods in other domains of physics  not related to four dimensional gauge theories and improving our understanding of the strong interaction and hadron physics due to other non-perturbative techniques that were not discussed in the book. Let us now briefly
fantasize  on hypothetical developments in those three avenues.
\subsection {Further progress in the application of the methods discussed in the book}
\begin{itemize}
\item
A lesson that follows from the book is that the exploration of physical systems on one space dimension is both simpler to handle and sheds light on the real world so there are plenty of other unresolved questions that could be explored first in two dimensions. This includes exploration of the full standard model and the physics beyond the standard model including supersymmetry and its dynamical breaking, large extra dimensions, compositeness etc.
\item
There has been a tremendous development in recent years in applying methods of integrable models and in particular of spin chains, like the thermal Bethe ansatz, to ${\cal N}=4$ SYM theory, namely, in the context of supersymmetric conformal gauge theory.  We have no doubt that there will be further development in computing the anomalous dimensions of gauge invariant operators and correlators.
\item
Moreover,  one can identify  in a similar manner to ${\cal N}=4$ SYM theory  a spin chain structure  in gauge theories which are confining and with less or even no supersymmetries. In that case the spin chain Hamiltonian would not correspond to the dilatation operator but rather be associated with the excitation energies of hadrons.
\item
It is plausible that the full role of magnetic monopoles and of instantons has not yet been revealed. They have already had several reincarnations and there may be more. For instance there was recently a proposal to describe baryons as instantons which are solitons of a five dimensional flavor gauge theory in curved five dimensions.

\subsection{ Applications to other domains}
\item
 A very important application of two dimensional conformal symmetry has been  to superstring theories. A great part of the developments in supserstring theories is attributed to the infinite dimensional conformal symmetry algebra.  In fact it went in both directions and certain progress in understanding the structure of conformal invariance has emerged from the research of string theories.
A similar symbiotic evolution took place with regard to the affine
Lie algebras.
\item
String theories and in particular the string theory on $AdS_5\times S^5$ have recently been analyzed using the tools of integrable models like mapping to spin chains, using the Behte ansatz equations, identifying a set of infinitely many conserved charges and using structure of Yangian symmetry.
\item
Spin chain models have been suggested to describe systems of ``real"
spins in condensed matter physics. As is discussed in this book the
application of the corresponding tools to field theory systems has
been quite fruitful. The opposite direction will presumably also
take place and the use of properties of integrability that were
understood in field theories will shed new light on certain
condensed matter systems.
\item
The application of conformal invariance to condensed matter systems
at criticality has a long history. There has been recently an
intensive effort to further develop the understanding of systems
like various superconductors, fractional Hall effect and other
systems using modern conformal symmetry techniques.

\subsection{ Developments in gauge dynamics due to other methods}
\item
An extremely important framework for  analyzing gauge theories  has
been  supersymmetry. Regardless  of whether  it is realized in nature or not
it is evident that there are more tools to handle  supersymmetric
gauge theories and hence they are much better understood than non
supersymmetric ones.  One can gain novel insight into non
supersymmetric theories by introducing supersymmetry breaking terms
to well understood supersymmetric models. For instance one can start
with the Seiberg-Witten solution of ${\cal N}=2$  where the
structure of vacua is known and extract confinement behavior in
${\cal N}=1$ and non supersymmetric theories.
\item
A breakthrough in the understanding of gauge theories in the strong
coupling regime took place with  the discovery by Maldacena of the
AdS/CFT holographic duality [160]. The strongly coupled ${\cal N}=4$
in the large $N_c$ limit and large  't Hooft parameter $\lambda$ is mapped into a
weakly curved supergravity background. Thousands of research papers
that followed develop this map in many different directions and in
particular also in relation to the pure YM theory and QCD in four
dimensions. There is very little doubt that further exploration of
the duality will shed new light on QCD and on hadron physics.

\item
String theory had been born as a possible theory  of hadron physics.
It then underwent a phase transition into a candidate of the theory
of quantum gravity and even a unifying theory of everything. In
recent years, mainly due to the AdS/CFT duality there is a
renaissance of the idea that hadrons at low energies should be
described as strings. This presumably combined with the duality
seems to be a useful tool that will improve our understanding of
gauge dynamics.
\item
The computations of scattering amplitudes in gauge theories has been boosted in recent years due to various developments including  the use of techniques based on twistors, on a novel T-duality in the context of the Ads/CFT duality
 and on a conjectured  duality between Wilson lines and scattering amplitudes. One does not need a wild imagination to foresee a further progress in the industry of computing scattering amplitudes.

\end{itemize}
To summarize, non-perturbative methods have always been  very important tools in exploring the physical world. We have no doubt that they will continue to be a very essential ingredient  in future developments of science in general and  physics in particular.
\newpage

\section{The list of references of the book}

\vskip 4cm

\section{ Appendix A- The table of content of the book }

\begin{figure}[!]
\begin{center}
\includegraphics[width=170mm]{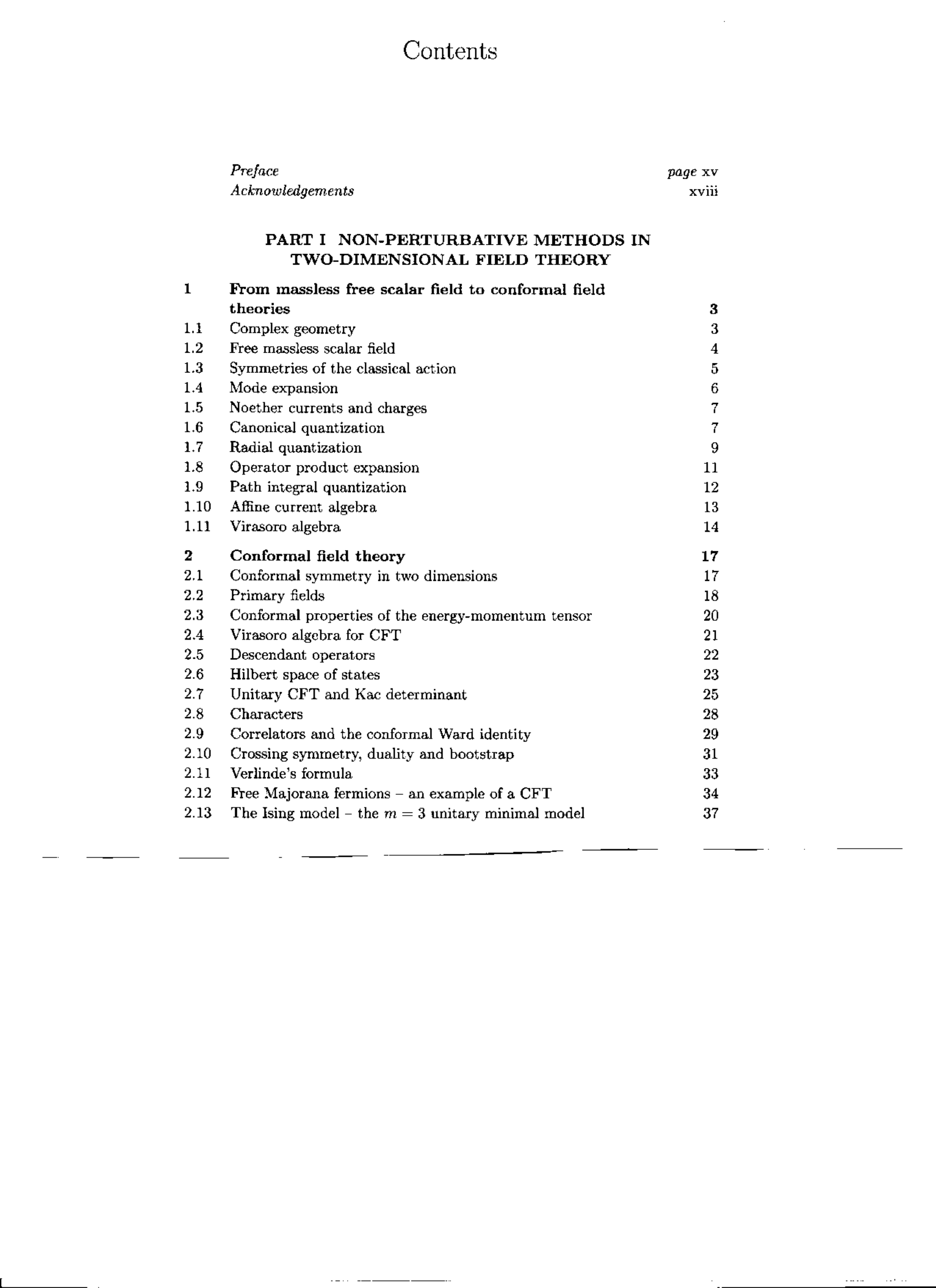}\end{center}
\end{figure}

\begin{figure}[!]
\begin{center}
\includegraphics[width= 170mm]{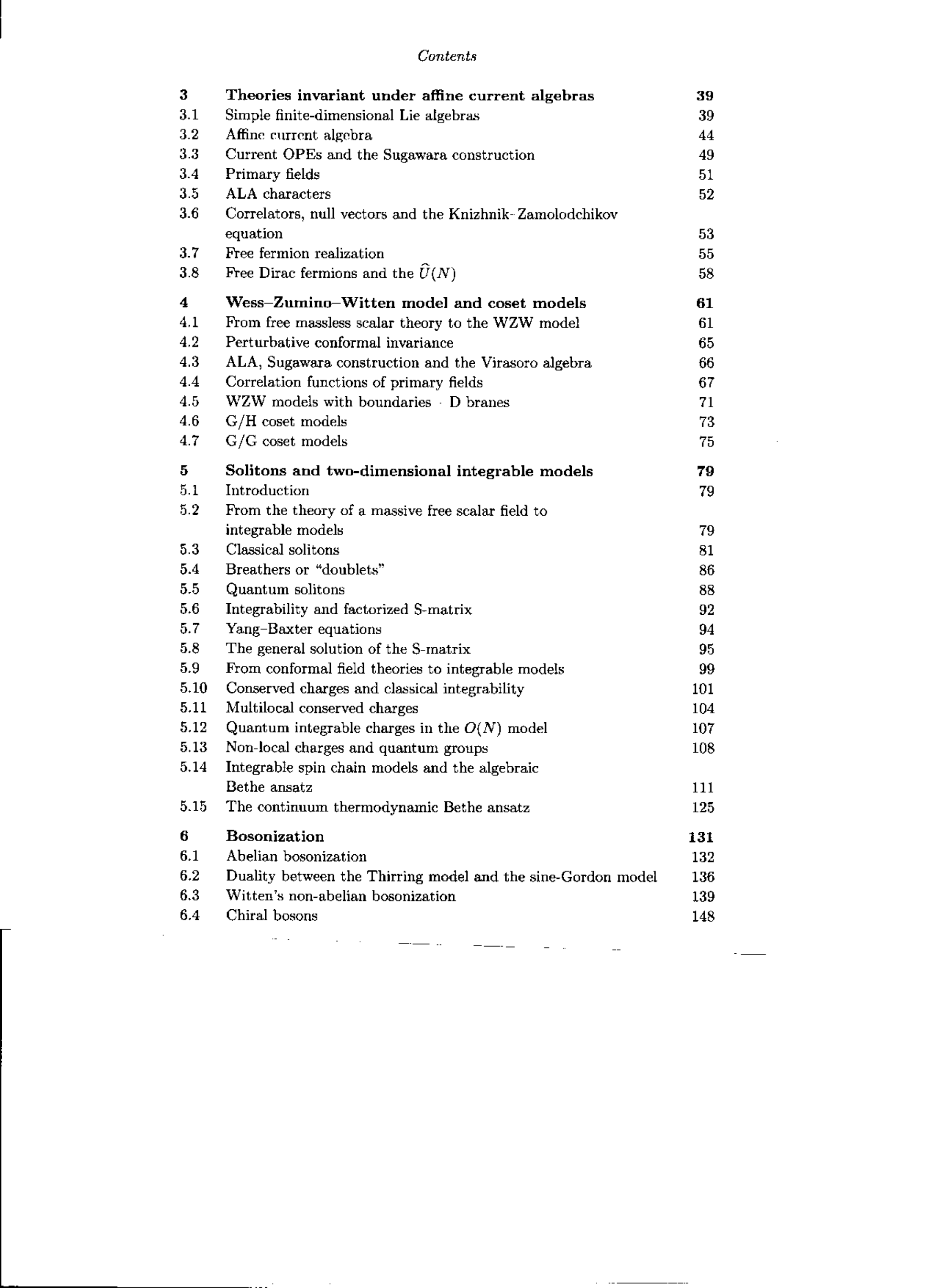}\end{center}
\end{figure}

\begin{figure}[!]
\begin{center}
\includegraphics[width= 170mm]{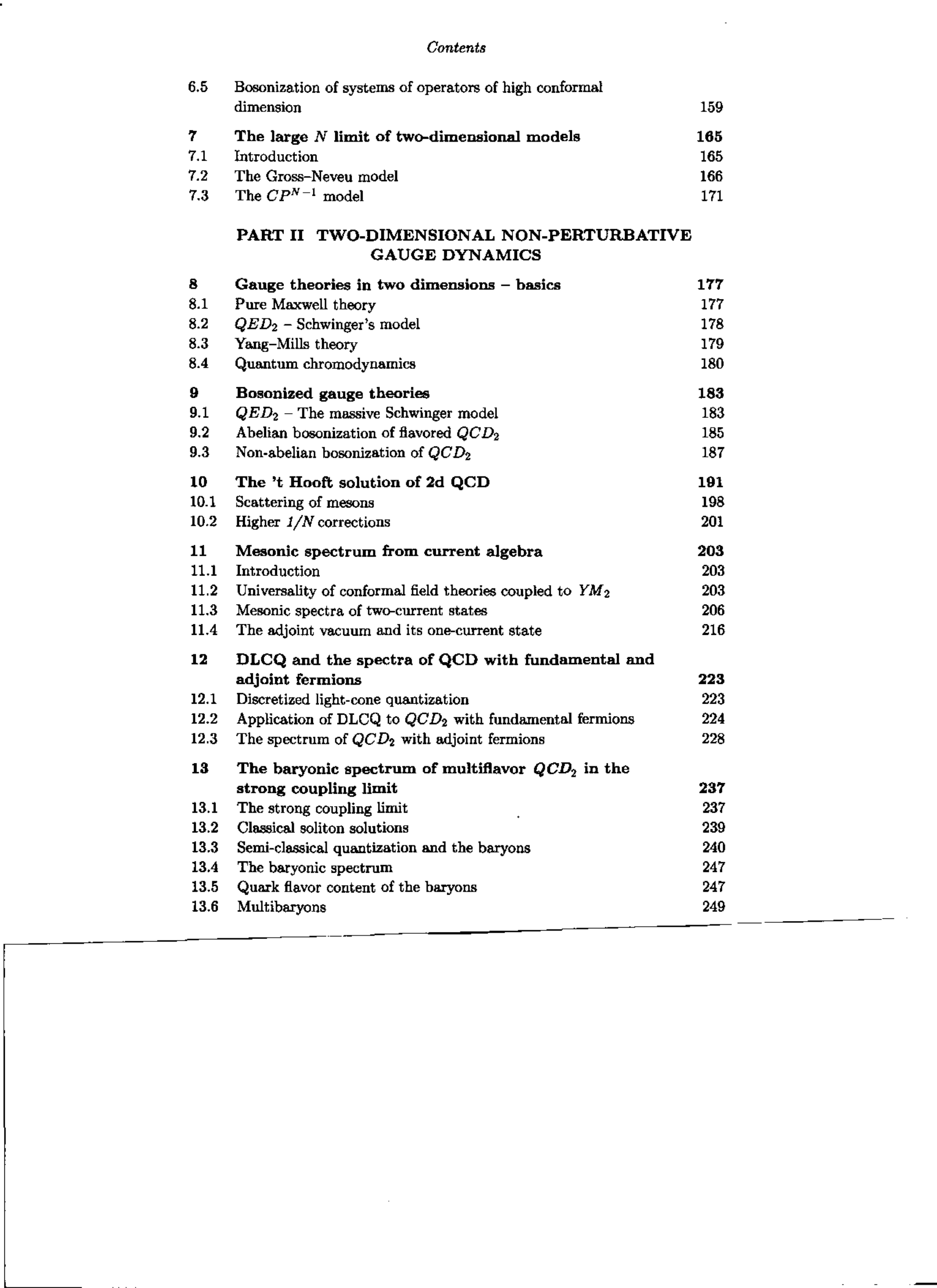}\end{center}
\end{figure}

\begin{figure}[!]
\begin{center}
\includegraphics[width= 170mm]{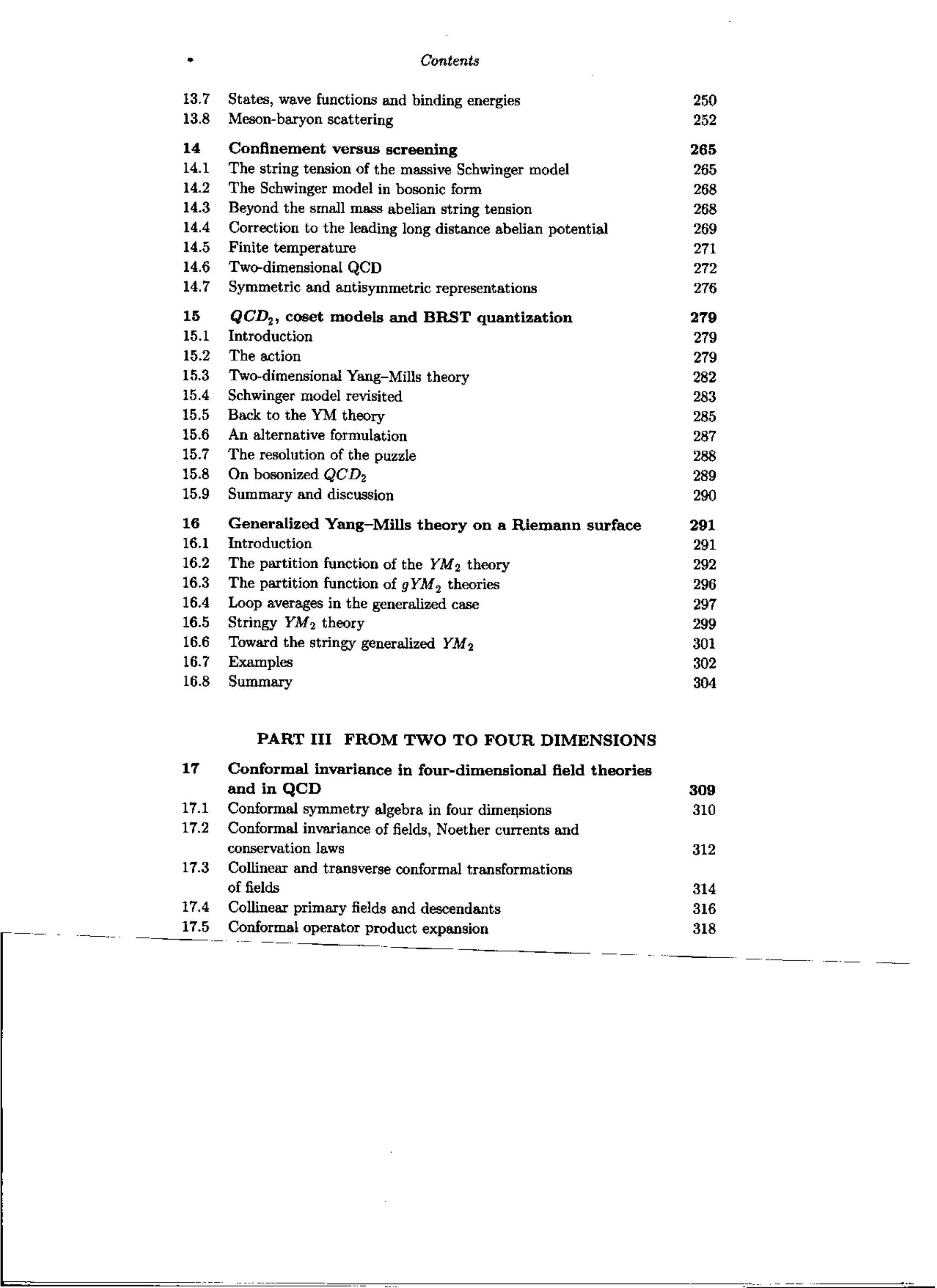}\end{center}
\end{figure}

\begin{figure}[!]
\begin{center}
\includegraphics[width= 170mm]{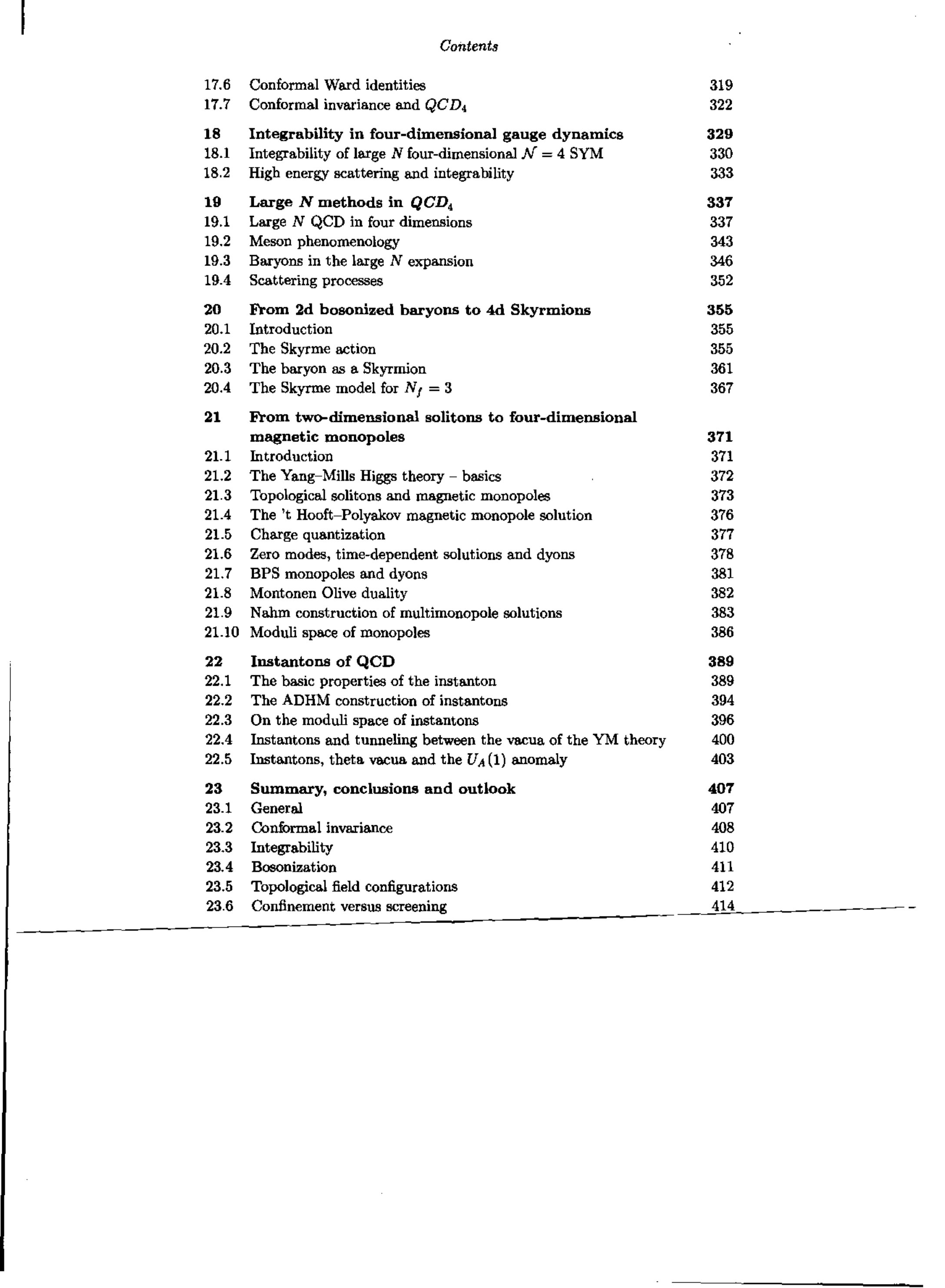}\end{center}
\end{figure}

\begin{figure}[!]
\begin{center}
\includegraphics[width= 170mm]{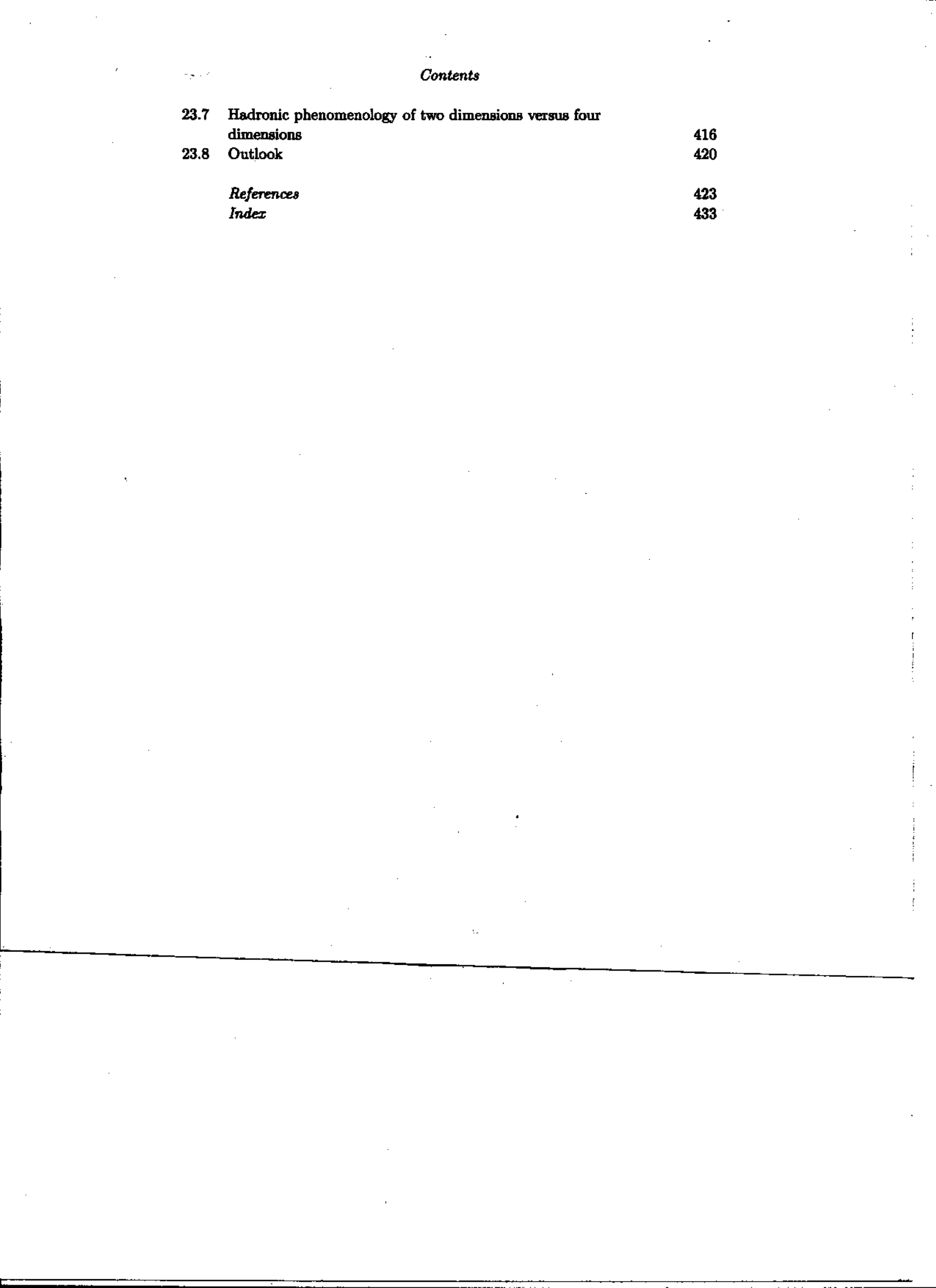}\end{center}
\end{figure}

\end{document}